\documentclass[preprint2]{aastex}
\newcommand{\etal}{{et \it al.}}
\newcommand{\hrs} {$^{\rm h}$}
\newcommand{\mins}{$^{\rm m}$}

\slugcomment{DRAFT v2000.11.01}			% VERSION #

\begin{document}

\title{HST-NICMOS Observations of M31's Metal Rich Globular Clusters
and Their Surrounding Fields\altaffilmark{1} II. Results}

\author{Andrew W. Stephens \& Jay A. Frogel}
\affil{The Ohio State University, Department of Astronomy}
\affil{140 West 18th Avenue, Columbus, OH  43210}

\author{Wendy Freedman \& Carme Gallart}
\affil{Carnegie Observatories}

\author{Pascale Jablonka}
\affil{Observatoire de Paris-Meudon}

\author{Sergio Ortolani}
\affil{Universit\`a di Padova}

\author{Alvio Renzini}
\affil{European Southern Observatory}

\author{R. Michael Rich}
\affil{University of California at Los Angeles}

\and 

\author{Roger Davies}
\affil{University of Durham}

\altaffiltext{1}{Based on observations with the NASA/ESA Hubble Space Telescope 
obtained at the Space Telescope Science Institute, which is operated by
AURA for NASA under contract NAS5-26555.}

\begin{abstract}
We have obtained HST-NICMOS observations of five of M31's most metal
rich globular clusters: G1, G170, G174, G177 \& G280.  For the two
clusters farthest from the nucleus we statistically subtract the field
population and estimate metallicities using $K$-$(J-K)$ color-magnitude
diagrams (CMDs).  Based on the slopes of their infrared giant branches
we estimate [Fe/H] $=-1.22\pm0.43$ for G1 and $-0.15\pm0.37$ for G280.
We combine our infrared observations of G1 with two epochs of optical
HST-WFPC2 $V$-band data and identify at least one LPV based on color and
variability.  The location of G1's giant branch in the $K$-$(V-K)$ CMD
is very similar to that of M107, indicating a higher metallicity than
our purely infrared CMD: [Fe/H]$\sim -0.9\pm0.2$.  For the three central
clusters, which are too compact for accurate cluster star measurements,
we present integrated cluster magnitudes and field CMDs.

The $K$-band luminosity functions (LFs) of the upper few magnitudes of
G1 and G280, as well as for the fields surrounding all clusters, are
indistinguishable from the LF measured in the bulge of our Galaxy.  This
indicates that these clusters are very similar to Galactic clusters, and
at least in the surrounding fields observed, there are no significant
populations of young luminous stars.

For the field surrounding G280, we estimate the metallicity to be $-1.3$
from the slope of the giant branch, with a spread of $\sigma_{[Fe/H]}
\sim 0.5$ from the width of the giant branch.  Based on the numbers and
luminosities of the brightest giants, we conclude that only a small
fraction of the stars in this field could be as young as 2 Gyr, while
the majority have ages closer to 10 Gyr.

\end{abstract}


%
% INTRODUCTION
%
\section{Introduction} \label{sec:introduction}

Globular clusters (GC) occupy a very special position in modern
astrophysics.  They provide the most stringent tests of stellar
evolution theory, represent ideal templates for stellar population
synthesis studies, allow age dating of galaxies with unrivaled
precision, etc.  The GC families of the Milky Way and its satellite
galaxies, the Magellanic Clouds and the Fornax dSph galaxy, have been
thoroughly studied from both the ground and space, especially with HST.
The next nearest major GC family belongs to M31.  To study individual
stars in GC systems much more distant than M31 will require the next
generation of ground or space based telescopes.

The GC systems of the the Milky Way and M31 appear to be quite similar
in luminosity and color ranges \citep{BBBF1987, EW1988, HBK1991,
FPC1980b, DGMC1997, SBA1996}.  However, some of M31's bulge globulars
appear to have line strengths as strong as those of giant ellipticals,
suggesting metallicities considerably greater than any known Galactic
globular \citep{BJSA1992, JAB1992, Jab1997}. \citet{BFGK1984} has also
noted that M31's GCs appear to follow different H$\beta$ and CN
correlations with Mg$_2$ with respect to Galactic globulars; they
attributed this to the whole M31 GC family being systematically younger.
Renzini (1986) pointed out that other interpretations were also
possible, such as the chemical enrichment history of the two spheroids
having proceeded with slightly different time scales, resulting in
different element ratios.

HST observations of M31's GCs in the optical (with the goal of stellar
photometry) started soon after the first refurbishing mission, with both
FOC and WFPC2 \citep{FBCC1996, RMFN1996, AGLB1996}, and now include some
of the most metal rich clusters \citep{JCMS2000}.  The study of metal
rich M31 globulars complements similar studies of Galactic bulge
globulars within this still poorly known, yet crucial part of
age-metallicity space.  Near-IR observations are essential to study the
brightest giants in an old, metal rich population, especially to
determine bolometric luminosities.  Metal rich globular and bulge
giants, which are the brightest bolometrically and in the near-IR, are
in contrast, many magnitudes fainter at optical wavelengths due to
severe molecular blanketing.  In extreme cases, $(V-K)$ can be as great
as $\sim 10$ at the top of the AGB \citep{FW1987, GOMR1998} and these
stars are likely to have escaped detection in M31 even with WFPC2.

Since the main sequences of the old populations in M31 are currently out
of reach, one is forced to appeal to an alternative age indicator.
Theory predicts that the highest luminosity reached on the AGB is a
function of age \citep{IR1983}, a prediction extensively verified by
observations of clusters in the Magellanic Clouds \citep{MA1986,
FMB1990}.  In metal poor Galactic globulars, no stars are brighter than
the theoretical RGB tip, as expected for a $\sim 14$ Gyr population.
However, the AGB of more metal rich clusters ([Fe/H]$>-1$) extends $\sim
1$ magnitude above the RGB \citep{FPC1983, FE1988, GRO1997}, as it does
in the Galactic bulge \citep{FW1987}.  For clusters, though, these
bright stars are all LPVs.  For both clusters and bulge population these
luminous stars are now generally ascribed to high metallicity, rather
than to young age \citep{FW1987, GRO1997}.  Hence, the presence of stars
brighter than the RGB tip does not guarantee an intermediate age;
consideration of their color, luminosity {\it and} frequency in the
parent population is required before drawing conclusions about ages
\citep{Ren1993}.

In Cycle 7 we proposed to obtain NICMOS $JHK$ images of 5 metal rich
globular clusters in M31.  From these observations, we planned to
achieve the following scientific goals: (1) explore the upper end of the
GC luminosity function; (2) derive independent estimates for the
metallicity of the M31 clusters and their adjacent fields from the slope
and location of the NIR RGB \citep{KFTP1995}; (3) determine the
frequency of luminous RGB and AGB stars (including LPVs) per unit
luminosity; (4) compare the cluster results with observations of M31
field stars (5) compare the properties of the luminous stars in M31's
metal rich clusters to those in the Galactic bulge and Galactic
globulars; (6) integrate these results with optical photometry and
spectroscopy and explore implications for stellar evolution theory and
the interpretation of the integrated light of distant galaxies.

One critical issue deserves special attention if any of these goals are
to be attained: the effect of stellar crowding.  We \citep[][hereafter
Paper I]{SFFJ2001}, have carefully analyzed the effects of blending on
our NICMOS data.  Through the creation of completely artificial
clusters, we have calculated threshold- and critical-blending limits for
each cluster and surrounding field.  These limits determine the
proximity to each cluster where reliable photometry can be obtained.
These simulations allow us to quantify and correct for the effects of
blending on the GB slope and width at different surface brightness
levels.

This paper is organized as follows.  Section \ref{sec:observations}
presents the reasons for selecting each cluster, and the details of the
observations.  Section \ref{sec:data_reduction} describes the reduction
procedures, and gives a brief summary of the procedures and results on
blending from Paper I.  Section \ref{sec:photometry} presents the
integrated photometry of the clusters, the CMDs and luminosity functions
of the clusters and their surrounding fields, and metallicity estimates
for G1, G280, and the G280 field.  Our conclusions are summarized in
Section \ref{sec:conclusions}.

%
% OBSERVATIONS
%
\section{Observations} \label{sec:observations}

We have obtained HST NICMOS images of five of M31's metal rich globular
clusters and their surrounding fields (Cycle 7; Program ID 7826).  These
observations are summarized in Table \ref{tab:observations}.  Column (1)
lists the ID of \citet{SKHV1977}, columns (2) and (3) give the center of
each field observed with HST, and column (4) lists the angular
separation from the nucleus of M31.  Columns (5-8) give previous
metallicity estimates for each cluster based on: (5) absorption
strengths from integrated optical spectra \citep{HBK1991}; (6)
spectroscopy of metal lines \citep{JAB1992,Jab1997}; (7) RGB
morphology \citep{FBCC1996,RMFN1996,Jab1997}; (8) integrated
ground-based NIR colors \citep{FPC1980b,CM1994}.  The last column (9)
gives the observation date of each target.

\begin{deluxetable}{ccccccccc}
\tablewidth{13cm}
\tablecaption{M31 Globular cluster observations}
\tabletypesize{\footnotesize}
\tablehead{
\colhead{ID \tablenotemark{a}}	&
\colhead{$\alpha$}	 	&
\colhead{$\delta$} 		&
\colhead{r}			&
\multicolumn{4}{c}{\hspace{2cm}[Fe/H]\hspace{-3cm}\rule[-0.15cm]{4.9cm}{0.02cm}} &
\colhead{Date}			\\
\colhead{}			&
\colhead{(2000)}		&
\colhead{(2000)}		&
\colhead{($'$)}			&
\colhead{HBK}			&
\colhead{JAB}			&
\colhead{F-P}			&
\colhead{NIR}			&
\colhead{UT}			}
\startdata
G001 & 00\hrs 32\mins 47\fs2 & 39\degr $34' 48''$ & 152.3 & -1.08 &-0.52   & -0.60(?)& -1.38 & 1998.07.18 \\
G170 & 00\hrs 42\mins 32\fs4 & 41\degr $10' 29''$ &   6.1 & -0.31 & 0.20   &  high   &\nodata& 1998.08.10 \\
G174 & 00\hrs 42\mins 33\fs3 & 41\degr $17' 17''$ &   2.6 &  0.29 & $>0.5$ & \nodata & -0.13 & 1998.08.13 \\
G177 & 00\hrs 42\mins 34\fs4 & 41\degr $14' 04''$ &   3.2 & -0.15 & 0.52   &  high   & -0.32 & 1998.09.08 \\
G280 & 00\hrs 44\mins 29\fs5 & 41\degr $21' 36''$ &  20.5 & -0.70 & \nodata& -0.40(?)& -0.40 & 1998.09.13 
\enddata
\tablenotetext{a}{\citet{SKHV1977}}
\label{tab:observations}
\end{deluxetable}

G174 is the most metal rich cluster known in M31 \citep{HBK1991,
Jab1997}.  It and G177 have spectra with line strengths comparable to
those of strong lined ellipticals, and stronger than the most metal rich
Galactic globulars.  G170's lines are somewhat weaker, but still
comparable to those of two of the most metal rich Galactic globulars,
NGC6528 and NGC6553 \citep{BROB1999, CGBC1999}.  G170, G174 and G177 are
projected very close to the nucleus of M31.  The G177 field, although
not the closest to M31's nucleus, lies along the major axis of M31 and
has the highest number of detected stars, $\sim 7$ arcsec$^{-2}$, as
well as the highest background, $\mu_K \sim 15.4$ magnitudes
arcsecond$^{-2}$.

G280's CMD has been characterized by \citet{FBCC1996} as similar to, but
not quite as metal rich, as NGC6553, consistent with its near-IR colors
\citep{FPC1980b}.  In spite of G280's similarity to G170, it lies much
farther from the nucleus of M31, at a distance of $20.5'$.  G280 is also
one of the most ``open'' clusters, making it one of the best clusters in
our sample for individual stellar photometry.

Finally, G1 is the largest, brightest globular cluster in M31.  This
cluster is of particular interest because of its high metallicity
despite its large distance ($152'$) from the center of M31.  Since it
nearly fills a NIC2 frame, we also observed a nearby field to estimate
the background stellar contribution.  It turns out that G1 is far enough
from M31's center that field star contamination is negligible, and our
control field yielded only 2 stars over the entire $20''$ dithered
$K$-band field.

Our observations were taken with the NICMOS camera 2 (NIC2) which has a
plate scale of $\sim$ 0\farcs0757 pixel$^{-1}$ and a field of view of
19\farcs4 on a side (376 arcsec$^2$).  The NICMOS focus was set at the
compromise position 1-2, which optimizes the focus for simultaneous
observations with cameras 1 and 2.  All of our observations used the
{\sc multiaccum} mode \citep{Mac1997} because of its optimization of the
detector's dynamic range and cosmic ray rejection.  Observations with
NIC1 will be described in a later paper concentrating on M31's bulge.

Each of our targets was observed through three filters: F110W (0.8--1.4
\micron), F160W (1.4--1.8 \micron), and F222M (2.15--2.30 \micron).
These filters are close to the standard ground-based $J$, $H$, \& $K$
filters.  The observation of each cluster spanned three orbits of HST,
with $\sim 42$ minutes of observing per orbit.  This yielded total
integration times of 1920s in F110W, 3328s in F160W, and 2304s in F222M
(see Table 2).

We implemented a spiral dither pattern with 4 positions to compensate
for imperfections in the infrared array.  The dither steps were 0\farcs4
for the $J$ and $K$ band images, and 5\farcs0 for the $H$ band images.
Thus the combined dithered images are $\sim 20''$ in $J$ and $K$, and
$\sim 24''$ in $H$.  We used the predefined sample sequences {\sc
step32} with 22 samples in $J$ and 25 samples in $K$, and the {\sc
step64} sequence with 21 samples in $H$.

We present $H$-band images of each cluster in Figures 1a-e.  These
images are the combination of 4 dither positions, and are $\sim 24''$ on
a side.  The $H$-band images are the deepest and also cover the most
area as they were acquired using the largest dithers.  Thus the $H$-band
provides the deepest luminosity function, and gives us additional color
information for LPV identification.  The other analyses use primarily
the $J$- and $K$-bands, as they are the bands where the groundbased
comparisons, age-luminosity relations, and metallicity indicators exist.

% Figure 1a
\begin{figure}
\plotone{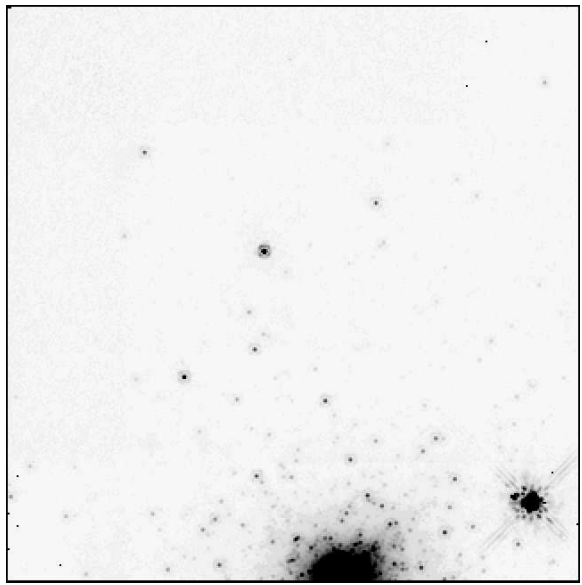}
\figcaption{
G1 -- F160W ($H$-band) combination of all 4 dithers; 3328s total
exposure.  The faintest stars seen in this image have $H \sim 21.8$
magnitudes.  The object in the lower right corner is a foreground star
and gives an idea of the full NICMOS PSF.
\label{fig:clusters}}
\end{figure}

% Figure 1b
\begin{figure}
\figurenum{1b}
\plotone{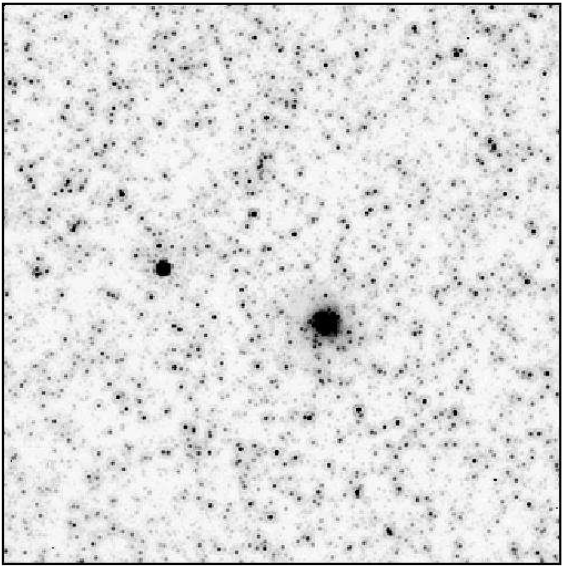}
\figcaption{
G170 -- Combination of 4 dithers in the F160W ($H$) filter, yielding
3328s total exposure time.}
\end{figure}

% Figure 1c
\begin{figure}
\figurenum{1c}
\plotone{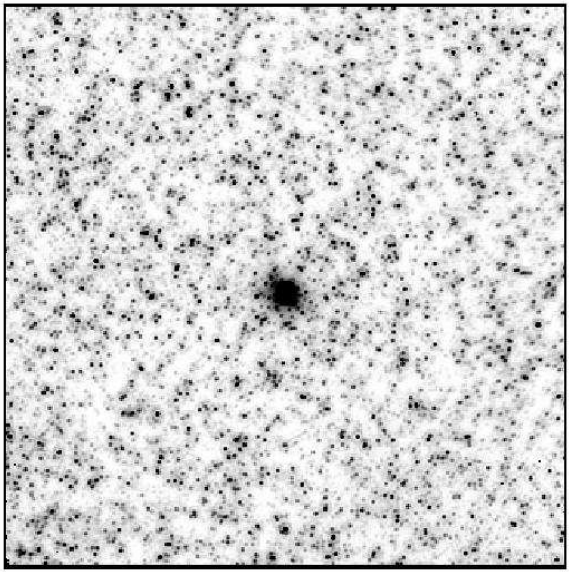}
\figcaption{G174 -- same as Fig. 1b}
\end{figure}

% Figure 1d
\begin{figure}
\figurenum{1d}
\plotone{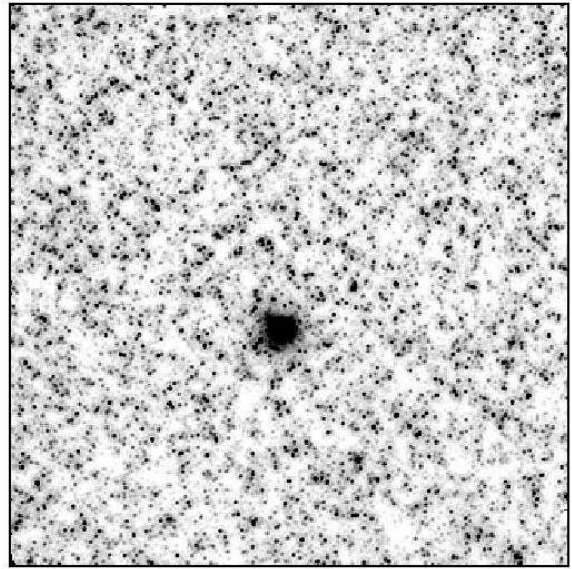}
\figcaption{G177 -- same as Fig. 1b}
\end{figure}

% Figure 1e
\begin{figure}
\figurenum{1e}
\plotone{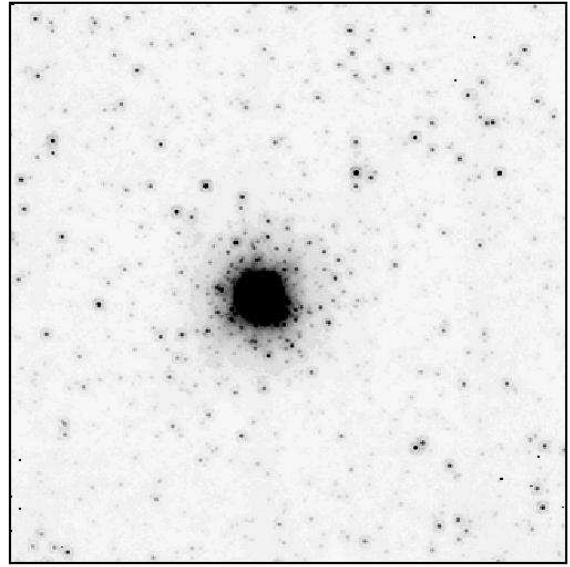}
\figcaption{G280 -- same as Fig. 1b}
\end{figure}

% Figure 1f
\begin{figure}
\figurenum{1f}
\plotone{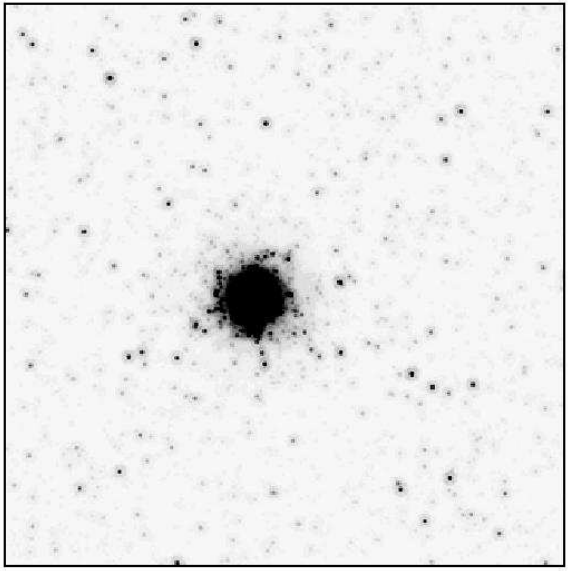}
\figcaption{
A280 -- The artificial analog of the G280 cluster and field.
Note the similarity to the real frame displayed in Fig. 1e.}
\end{figure}

%
% DATA REDUCTION
%
\section{Data Reduction} \label{sec:data_reduction}

Our data were reduced with the STScI pipeline supplemented by the IRAF
NICPROTO package (May 1999) to eliminate any residual bias (the
``pedestal'' effect).  Object detection was performed on a combined
image made up of all the dithers of all the bands (12 images in total).
PSFs were determined from each of the four dithers, then averaged
together to create a single PSF for each band of each target (the
average FWHM of each band is listed in Table \ref{tab:filters}).
Instrumental magnitudes were measured using the ALLFRAME PSF fitting
software package \citep{Ste1994}, which simultaneously fits PSFs to all
stars on all dithers.  DAOGROW \cite{Ste1990} was used to determine the
best magnitude in a $0.5''$ radius aperture, which we then converted to
the CIT/CTIO system using the transformation equations of
\citet{SFOD2000}.

\begin{figure}
\plotone{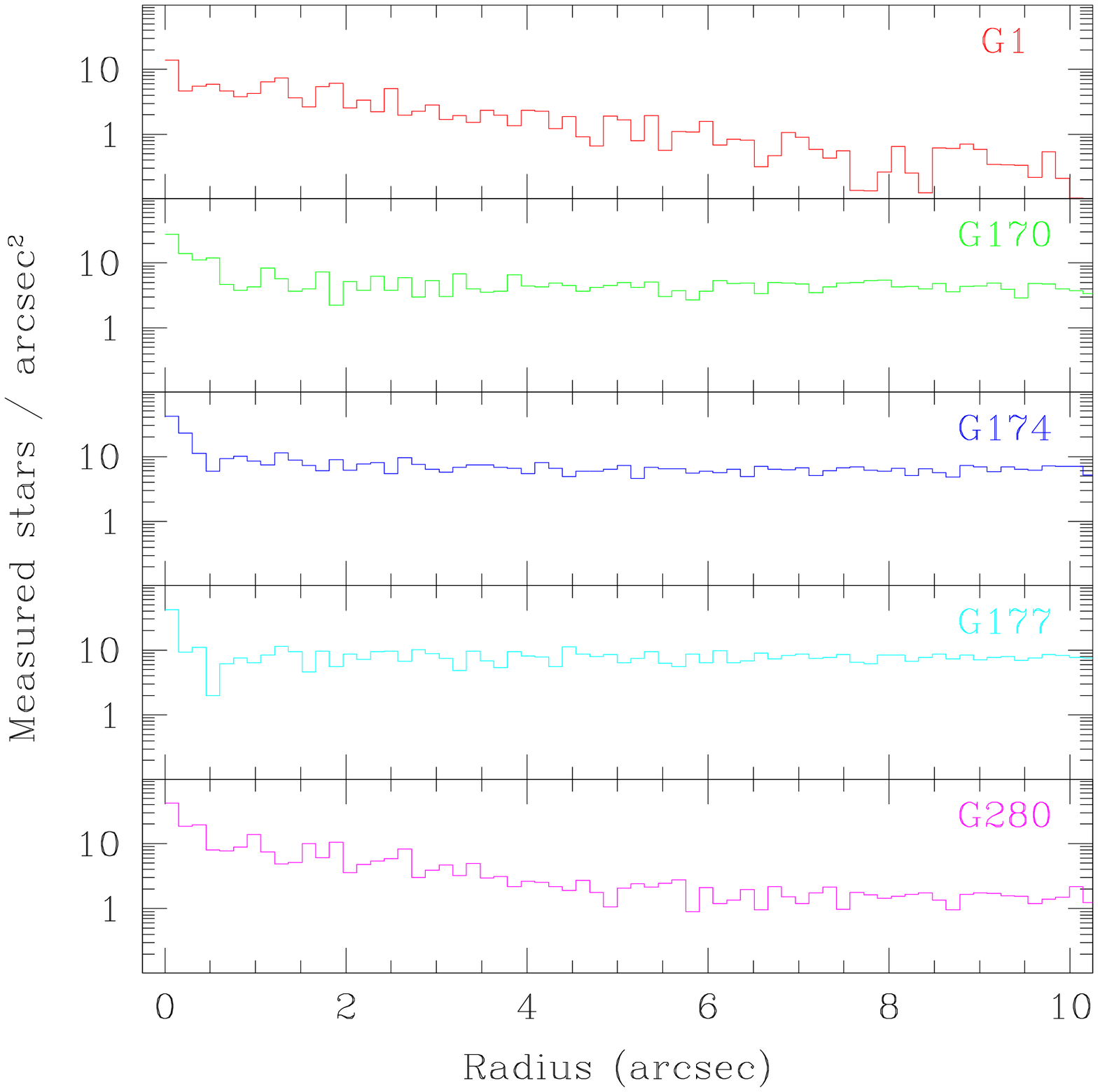}
\figcaption{
Number counts as a function of radius from the cluster center.
\label{fig:stellar_density}}
\end{figure}

The azimuthally averaged number of detected stars per square arcsecond
for each cluster is shown in Figure \ref{fig:stellar_density} as a
function of radius from the cluster center.  The counts include stars
which were measured at least once in any band.  Even though it appears
we have detected stars into the centers of all the clusters, we have
demonstrated in Paper I that the detections near the cluster cores are
spurious, the result of image blending.  The three clusters near the
center of M31, G170, G174 and G177, have central spikes in the number
counts, but quickly fade into the background.  No photometry is possible
for these clusters.  For the two less compact clusters, G1 and G280, the
number of detections decreases gradually with radius, and photometry is
possible in their central regions.

\begin{deluxetable}{cccc}
\tablewidth{5.5cm}
\tablecaption{NICMOS Filters}
\tabletypesize{\footnotesize}
\tablehead{
\colhead{Filter}	&
\colhead{Exposure} 	&
\multicolumn{2}{c}{FWHM} \\
\colhead{}		&
\colhead{(s)}		&
\colhead{(pix)}		&
\colhead{($''$)}	}
\startdata
F110W	& 1920	& 1.65	& 0.13	\\
F160W	& 3328	& 1.95	& 0.15	\\
F222M	& 2304	& 2.45	& 0.19	\\
\enddata
\label{tab:filters}
\end{deluxetable}

%
% Blending Analysis
%
\subsection{Blending Analysis} \label{sec:blending_analysis}

Stellar photometry in crowded regions such as we have observed can be
strongly affected by blending.  In order to quantify and attempt to
correct for blending, we have performed several types of artificial star
experiments; they are described in detail in Paper I.  For these
experiments we constructed completely artificial clusters.  Starting
with a blank frame having the appropriate noise characteristics, we
added stars according to the cluster's radial profile.  The input
stellar population was chosen to match the luminosity function and
colors of giants observed in the Galactic bulge.  The artificial frames
were then processed and measured in exactly the same manner as the real
data.  As an example, A280, the artificial analog of the G280 cluster,
is shown in Figure 1f.  This frame is composed of 450,000 cluster stars
and 80,000 field stars.  The CMDs and LFs from the artificial clusters
can be used to determine the origin of objects seen in the real CMD, and
the validity of our measured LFs.

To better study the effects of blending at different surface brightness
levels, we have also created uniformly populated mini-fields.
Constructed in the same manner as the artificial clusters, these $100
\times 100$ pixel fields each have between $10^4$ and $\sim 5 \times
10^6$ stars, enough to reach the surface brightnesses observed in the
cores of these M31 clusters.

Using Figure 9 of Paper I, which show the difference between the
recovered and input stellar magnitudes as a function of the field
surface brightness, we have chosen $\mu_K = 16$ magnitudes
arcsecond$^{-2}$ as the threshold surface brightness where our
photometry starts to become noticeably affected by blending, and $\mu_K
= 14$ mag arcsec$^{-2}$ as the critical surface brightness where our
photometry is dominated by blends, and no longer yields any useful
information.  Brighter than this level, no measurements are reliable.
If one desires to accurately measure stars fainter than those we have
measured, the threshold surface brightness will have to be fainter than
$\mu_K = 16$.

\begin{figure}
\plotone{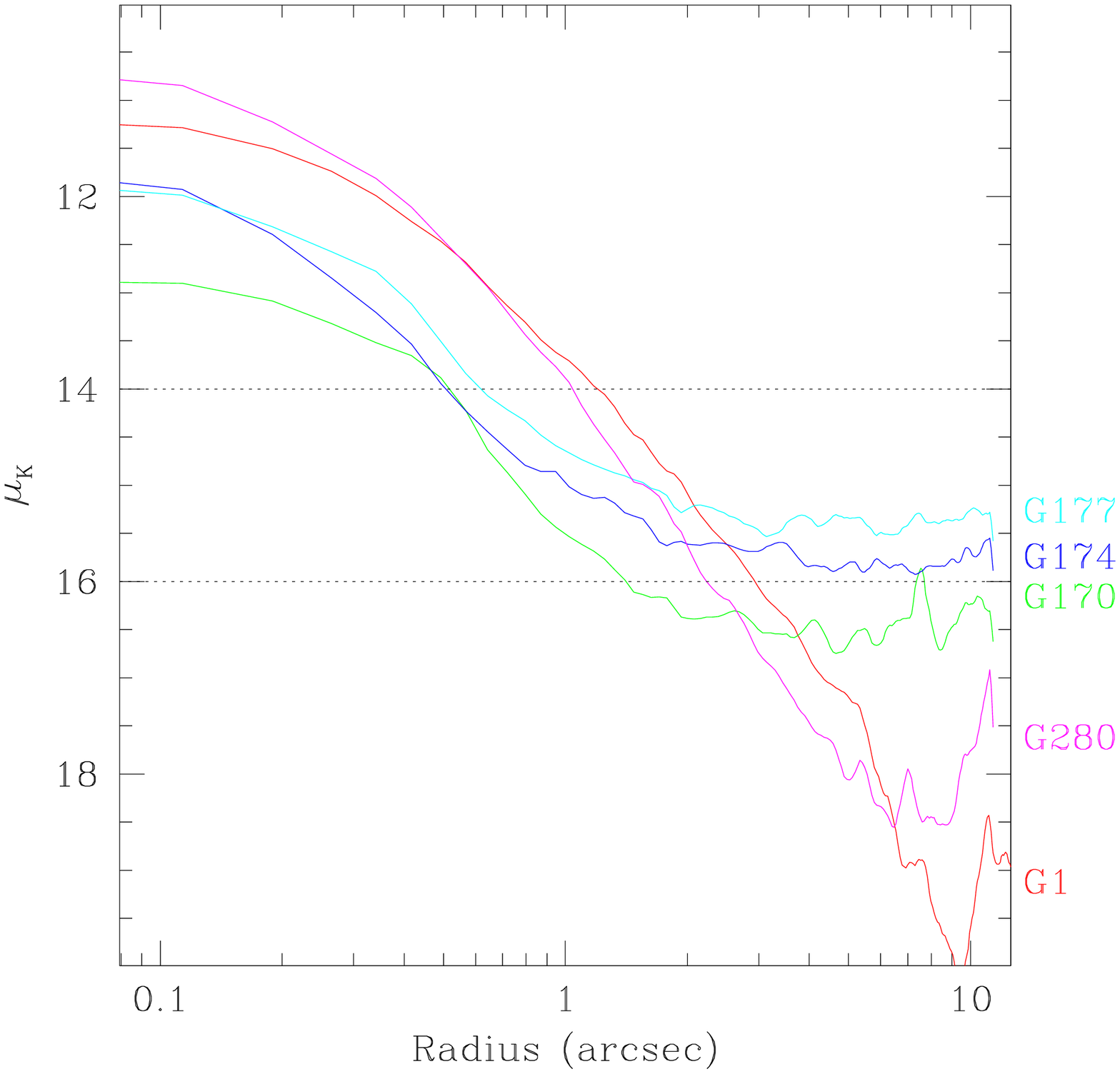}
\figcaption{
Azimuthally averaged $K$-band surface brightness as a function of radius
from the cluster center.
\label{fig:surface_brightness}}
\end{figure}

Figure \ref{fig:surface_brightness} shows azimuthally averaged $K$-band
surface brightness profiles of each cluster.  Dotted lines indicate our
chosen threshold- and critical-blending surface brightness levels.  We
use this plot to determine the threshold-blending radius ($R_{16}$), and
the critical-blending radius ($R_{14}$) of each cluster.  These radii
are listed in Table \ref{tab:blend_radii}.  Any objects measured inside
the threshold-blending radius, especially faint objects, are potentially
affected by blending, and should be considered suspect.  (Note that the
radius $R_{16}$ was chosen so that stars input at $M_K \sim -3$ could be
recovered accurately most of the time.  However, for stars fainter than
this, there is no way to tell whether stars measured at $M_K > -3$ are
blends or not.)  Objects measured inside the critical-blending radius
are undoubtedly blends, and although we plot them for completeness, they
should be disregarded.

Since we want to use the slope of the GB to estimate metallicities
\citep{KF1995}, and the width of the GB to place limits on any spread in
metallicity, we have investigated the effects of crowding on these
quantities.  At low surface brightnesses ($\mu_K \sim 20$ magnitudes
arcsecond$^{-2}$) blending has a negligible effect on the measurement of
the GB slope, and the correct metallicity is calculated.  As the surface
brightness increases to $\mu_K \sim 13$ magnitudes arcsecond$^{-2}$, the
recovered GB slope also increases (becomes more negative), yielding an
artificially greater metallicity.  Plotting the recovered GB slope as a
function of surface brightness from the artificial frames, we have
determined a linear correction to the slope in an attempt to account for
the effects of blending (Fig. 11, Paper I); we apply this relation to
our observed slope to improve our metallicity determinations.  The true
change in slope due to blending will, of course, depend upon the true
luminosity function and stellar colors.

\begin{deluxetable}{cccc}
\tablewidth{5cm}
\tablecaption{Threshold \& Critical Blend Radii}
\tabletypesize{\footnotesize}
\tablehead{
\colhead{Name}                  &
\colhead{Core $\mu_K$}       &
\colhead{$R_{14}$}              &
\colhead{$R_{16}$}              }
\startdata
G001	& 11.2 	& 1.2 	& 2.9		\\
G170	& 12.9 	& 0.5 	& 1.4 		\\
G174	& 11.7 	& 0.5 	& \nodata	\\
G177	& 11.8 	& 0.6 	& \nodata	\\
G280	& 10.7 	& 1.0 	& 2.2 
\enddata
\label{tab:blend_radii}
\end{deluxetable}

%
% PHOTOMETRY
%
\section{Photometry} \label{sec:photometry}

Here we present the integrated cluster photometry, and individual
stellar photometry for the clusters and fields.  Using the criteria
developed in paper I, and summarized in \S \ref{sec:blending_analysis}
here, we reject measurements which may be affected by blending.  As
stated above, cluster star measurements could only be extracted for the
two less compact clusters G1 and G280.  For these two clusters we
perform statistical subtraction of the field to yield a cluster GB.  Our
simple technique to do statistical subtraction divides the stars into
cluster ($2.2<r<5''$) or field ($r>5''$), then partitions each into bins
in $JK$ space, 1 magnitude wide in $J$ and $K$; we chose rather wide
bins to allow for spread due to blending.  The number of field stars per
bin is then normalized by multiplying by the ratio of the cluster to
field areas, and the appropriate number of suspected field stars are
randomly subtracted from each of the cluster star bins.  Using stars
brighter than $M_K = -3.7$ and $M_J = -2$ and throwing away $3\sigma$
outliers, we perform an iterative linear least-squares fit to the GB.
We then estimate the cluster metallicity using the GB slope technique
\citep{KFTP1995, KF1995}, and compare the M31 GBs with those measured
for Galactic clusters.  For the three central clusters, we present only
the surrounding field photometry cleaned of any potential blends.  The
data are presented in the form of $M_K$-$(J-K)$ diagrams, as well as
luminosity functions, and a $V$-$(V-K)$ diagram of G1 using optical
WFPC2 data.  All data assume $(m-M)=24.4$ and no reddening.

%
% Integrated Photometry
%
\subsection{Integrated Photometry} \label{sec:integrated_photometry}

Using simple aperture photometry, we have measured integrated magnitudes
for G170, G174, G177, and G280 clusters.  These measurements are listed
in Table \ref{tab:integrated_photometry}.  The three central clusters
(G170, G174 \& G177) are very compact, and an aperture of 40 pixel
($3.03''$) radius is chosen as the optimum compromise between the
maximum aperture size and best sky measurement.  G280 is more extended,
and we therefore use a 60 pixel ($4.54''$) radius aperture.  The sky was
measured as the average of an annulus around the clusters, using the
largest outer radius possible, typically 60 pixels wide, stopping just
short of the bright region at the bottom of all of the $K$-band frames.
The formal errors are $J \pm 0.03$, $H \pm 0.03$, and $K \pm 0.04$,
however, the measurements are very dependent on our sky level estimate.
Since we are resolving, but trying to average out, the background stars,
the sky estimate is sensitive to the size and location of the background
region.

Previous measurements are listed on the right side of Table
\ref{tab:integrated_photometry}.  \citet{FPC1980b} used single-channel
photometry with a $7.5''$ radius aperture.  The \citet{CM1994} and
\citet{BHBF2000} works used infrared arrays with $2.8''$ and $6''$
radius apertures respectively.  Due to the limited size of our array, we
were unable to match the large apertures of some of the previous
measurements.  However, our excellent resolution allows us to very
carefully place our measurement aperture and sky annulus (e.g. to avoid
the bright field star $\sim 7.4''$ from G170).  The good color agreement
but poor magnitude agreement with previous observations suggests a
measurement problem that is present in all bands, and thus cancels out
in the color determination.  The most likely cause is the difference in
the cluster and sky measurement regions between us and previous authors.

\begin{deluxetable}{ccccc|cccc}
\tablewidth{11.5cm}
\tablecaption{Integrated Photometry}
\tabletypesize{\footnotesize}
\tablehead{
\colhead{}		&
\multicolumn{4}{c}{ \underline{ \hspace*{1.3cm} NICMOS       \hspace*{1.3cm} } } &
\multicolumn{4}{c}{ \underline{ \hspace*{1.1cm} Ground-based \hspace*{1.1cm} } } \\
\colhead{Cluster}	&
\colhead{$K$} 		&
\colhead{$J-K$}		&
\colhead{$H-K$}		&
\colhead{ap}		&
\colhead{$K$}		&
\colhead{$J-K$}		&
\colhead{$H-K$}		&
\colhead{ap}		}
\startdata
G170	& 12.93	& 0.98	& 0.36	& 3.0	& 12.84  & 0.98  & \nodata  & 6.0 \tablenotemark{a} \\
G174	& 12.56	& 1.03	& 0.47	& 3.0	& 12.84	 & 1.00	 & \nodata  & 2.8 \tablenotemark{b} \\
G177	& 12.36 & 0.98	& 0.39	& 3.0	& 12.50	 & 0.92  & \nodata  & 2.8 \tablenotemark{b} \\
G280	& 10.91	& 0.89	& 0.33	& 4.5	& 11.06  & 0.88  & 0.15	    & 7.5 \tablenotemark{c} 
\enddata
\tablenotetext{a}{\citet{BHBF2000}}
\tablenotetext{b}{\citet{CM1994}}
\tablenotetext{c}{\citet{FPC1980b}}
\label{tab:integrated_photometry}
\end{deluxetable}

\begin{deluxetable}{ccccc}
\tablewidth{9cm}
\tablecaption{G1 \& G280 Metallicities}
\tabletypesize{\footnotesize}
\tablehead{
\colhead{Reference} & \colhead{G1} & \colhead{G280} }
\startdata
\citet{HC1977}		& $-0.3$	& $ 0.0$	\\
\citet{FPC1980b}	& $-1.22$ 	& $-0.19$	\\
\citet{BDFF1987}	& $-1.23$	& $-0.37$	\\
\citet{HCFJ1988}	& $-0.7$	& \nodata	\\
\citet{HBK1991}		& $-1.08$ 	& $-0.70$	\\
\citet{JAB1992}		& $-0.52$ 	& \nodata	\\
\citet{RMFN1996}	& $-0.8$	& \nodata	\\
\citet{FBCC1996}	& $-1.14$	& $-0.44$	\\
This work: $(J-K)$ CMD	& $-1.22\pm0.43$& $-0.15\pm0.37$\\
This work: $(V-K)$ CMD	& $-0.9\pm0.2$	& \nodata
\enddata
\label{tab:g1_g280_metallicities}
\end{deluxetable}

%
% G1
%
\subsection{G1} \label{sec:g1}

\begin{figure}
\plotone{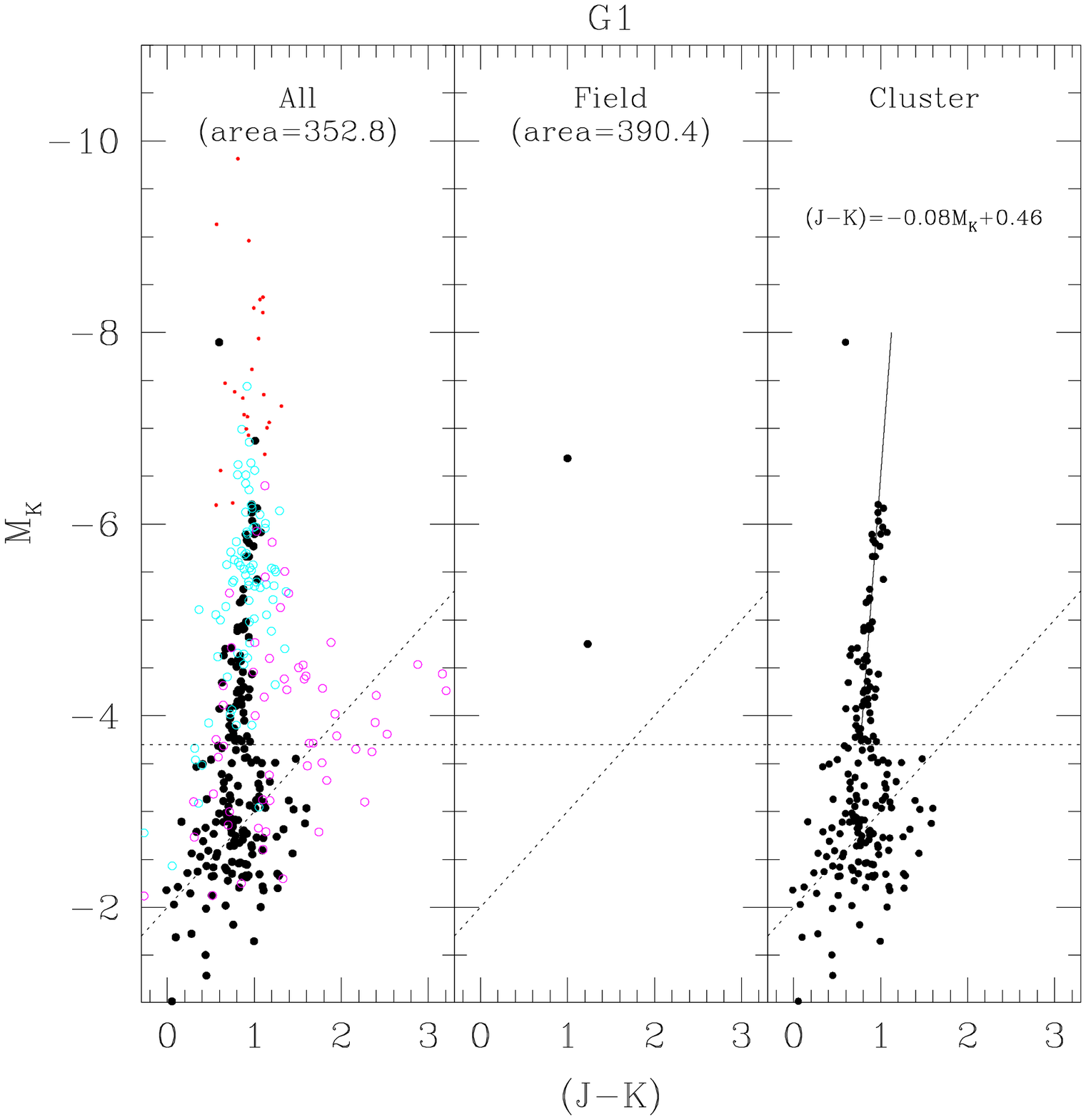}
\figcaption{
G1 -- Left: All objects measured on the G1 frame.  Open circles indicate
objects within the threshold-blending limit ($\mu_K<16, r<2.9''$) or
within 25 pixels of the bottom of the $K$-band image.  Half-size dots
are for objects inside the critical-blending limit ($\mu_K<14,
r<1.2''$).  Center: Stars measured in the adjacent control field $64''$
SE of G1.  Right: Cluster stars remaining after potential blends, noisy
measurements, and field stars have been removed.  The linear fit to the
GB above the 50\% completeness limits of $M_K=-3.7$ and $M_J=-2$ (dashed
lines) is shown at the top.
\label{fig:g1cmd}}
\end{figure}

The $M_K$-$(J-K)$ color magnitude diagram for the G1 cluster is shown in
Figure \ref{fig:g1cmd}.  The left panel shows all the cluster data from
the G1 frame.  Open circles indicate objects which are located inside
the threshold-blending radius ($2.9''$), or lie in a region of high
background in the lower 25 pixels of the the $K$-band frames (on the
CMD, these stars include all objects with $(J-K)>1.6$, and a similar
number of objects bluer and fainter).  Objects inside the
critical-blending radius ($1.2''$) are plotted with half-size dots.
The center panel shows the (two) objects measured in the adjacent
control field located $64''$ SE of G1.  The right panel shows only good
measurements of cluster stars.  Potentially blended objects and objects
within 25 pixels of the bottom of the frame have been removed, and (two)
field stars have been statistically subtracted.

We applied a linear least-squares fit to the cluster stars on the upper
GB, only using stars brighter than $M_K=-3.7$ and $M_J=-2$, the 50\%
completeness limits, and ignoring $3 \sigma$ outliers.  The best-fit
equation is displayed at the top of the right panel.  We then used the
relationship between the GB slope and globular cluster metallicity
derived by \citet{KFTP1995,KF1995} for Galactic globulars to estimate
the cluster metallicity.  This relation states that [Fe/H]$= -2.98 -
23.84 \times slope_{GB}$.  Our measured GB slope is $-0.083 \pm 0.014$,
which implies a metallicity of $-1.00 \pm 0.42$.  The error estimate is
a quadratic combination of the error in our linear fit, and the quoted
0.25 dex scatter observed in the GB slope -- metallicity relation.  Note
that the range in our fitted $M_K$ (3 mags) is significantly smaller
than the range (4.5 mags) \citet{KF1995} used to define the relation.

As discussed in \S \ref{sec:blending_analysis}, the measured slope of
the GB is affected by blending, even at relatively low surface
brightnesses.  Ideally we would calculate the GB slope in several
annuli, and correct each according the the average SB in that annulus.
In reality, there are so few stars that are measurable on the upper GB
to begin with, that splitting it up into even two annuli degrades the
accuracy of the slope determination significantly.  Thus we take a
number -- weighted average surface brightness of $\mu_K = 18.4$ for all
usable cluster photometry, which leads to a metallicity correction of
$-0.22 \pm 0.09$ dex, where the error of 0.09 dex is the rms scatter
around our correction.  This gives a final metallicity estimate of
$-1.22 \pm 0.43$ for G1.

\begin{figure}
\plotone{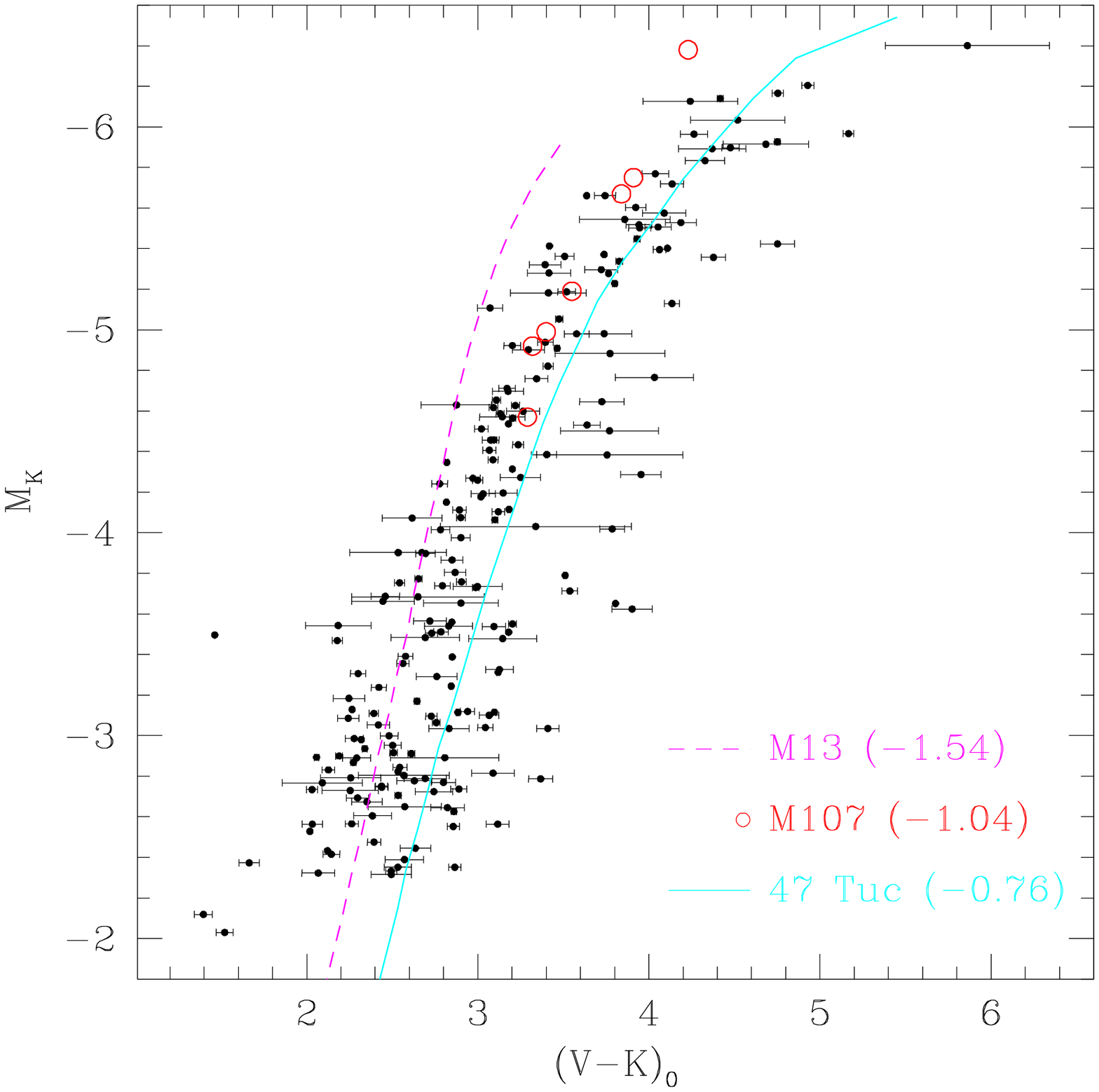}
\figcaption{
G1 CMD created through the combination of our $K$-band data with the
WFPC2 $V$-band observations of \citet{RMFN1996} and \citet{MSJ2000}.
The points are the mean $(V-K)$ from both optical datasets, and the
errorbars illustrate the difference between the two optical datasets.
Thus large errorbars indicate variability at the top of the CMD, and
most likely misidentifications at the bottom of the CMD.  G1 has been
corrected for distance and reddening with $(m-M) = 24.4$ and
$E(V-K)=0.28$ \citep{FPC1980b}.  Over-plotted are the giant branch ridge
lines of M13 (dashed line) and 47 Tuc (solid line) from \citet{FPC1981}.
Measurements of single stars in M107 from \citet{FPC1983} have also been
plotted for reference.  The metallicities (in parentheses), distances
and reddening for each cluster were taken from \citet{Har1996}.
\label{fig:g1_opticalcmd}}
\end{figure}

We have combined our infrared NICMOS data with the $V$-band WFPC2
observations of \citet{RMFN1996} (1994.07.29) and \citet{MSJ2000}
(1995.10.02).  The resulting $K$-$(V-K)$ CMD is shown in Figure
\ref{fig:g1_opticalcmd}.  In this diagram, the points indicate the mean
$(V-K)$ obtained from both optical datasets.  The errrorbars illustrate
the range of the observed $V$-band measurements.  Since the measurement
errors are relatively small, any large deviations are assumed to be
indicative of stellar variability.  Thus several of the most luminous
stars near the top of the GB are undoubtedly variables.  We have also
over-plotted the RGB ridge lines of M13 and 47 Tucanae from
\citet{FPC1981}, and stars in M107 from \citet{FPC1983}.  We point out
that the G1 RGB appears to lie blue-ward of 47 Tuc, and nearly on top of
the M107 measurements.  This indicates that G1 probably has a slightly
higher metallicity ([Fe/H]$\sim -0.9$) than we obtained from the slope of
the infrared GB.

Previous measurements of G1's metallicity are listed in Table
\ref{tab:g1_g280_metallicities}.  \citet{HC1977} used Washington
photometry of the integrated light to obtain a value of -0.3.
\citet{FPC1980b} used the integrated $(V-K)$ colors and the infrared CO
index, finding [Fe/H]$= -1.22$.  \citet{BDFF1987} recalibrated the
Frogel results using the new [Fe/H] values of \citet{Zin1984}, obtaining
[Fe/H]$= -1.23$.  Based on the GB position in the $I$-$(V-I)$ diagram,
\citet{HCFJ1988} estimated a metallicity of $\sim -0.7$, and \citet{RMFN1996} 
found a metallicity ``at least as high as 47 Tuc''.  Using the strengths
of absorption features in integrated optical spectra, \citet{HBK1991}
found a metallicity of $-1.08$.  Averaging the metallicities from many
metallic lines in the spectral range $\lambda \lambda 3200-9750$\AA,
\citet{JAB1992} determined $\log (Z/Z_0) = -0.52$ Most recently,
\citet{FBCC1996} using the $(V-K)_0$ color, and the calibration from
\citet{BH1990} found [Fe/H]$=-1.14$.

Our two metallicity determinations for G1 are different, but not
inconsistent.  We note that the value from the $(J-K)$ CMD ([Fe/H]$=
-1.22 \pm 0.43$) is based on a very small luminosity range, and has
quite large errors.  The estimate from the appearance of the $(V-K)$ CMD
([Fe/H]$\sim -0.9$) is less quantitative, but probably more robust.

There have been suggestions that the stars in G1 may have a range in
metallicity \citep{JBSM1999}, possibly due to self-enrichment.  If this
is so, we should be able to detect a spread in color in our near-IR
CMDs.  However, we find no evidence for such a spread as the dispersion
in either the $(J-H)$ or $(J-K)$ colors: $\sigma_{(J-H)}=0.06$ in the
range $-3 > M_H > -6$, and $\sigma_{(J-K)}=0.04$ in the range $-4.5 >
M_K > -6.5$.  These are both very close to the spread expected solely
from measurement errors, as predicted by the artificial cluster A1, and
by the ALLFRAME photometric uncertainties.

%
% G280
%
\subsection{G280} \label{sec:g280}

\begin{figure}
\plotone{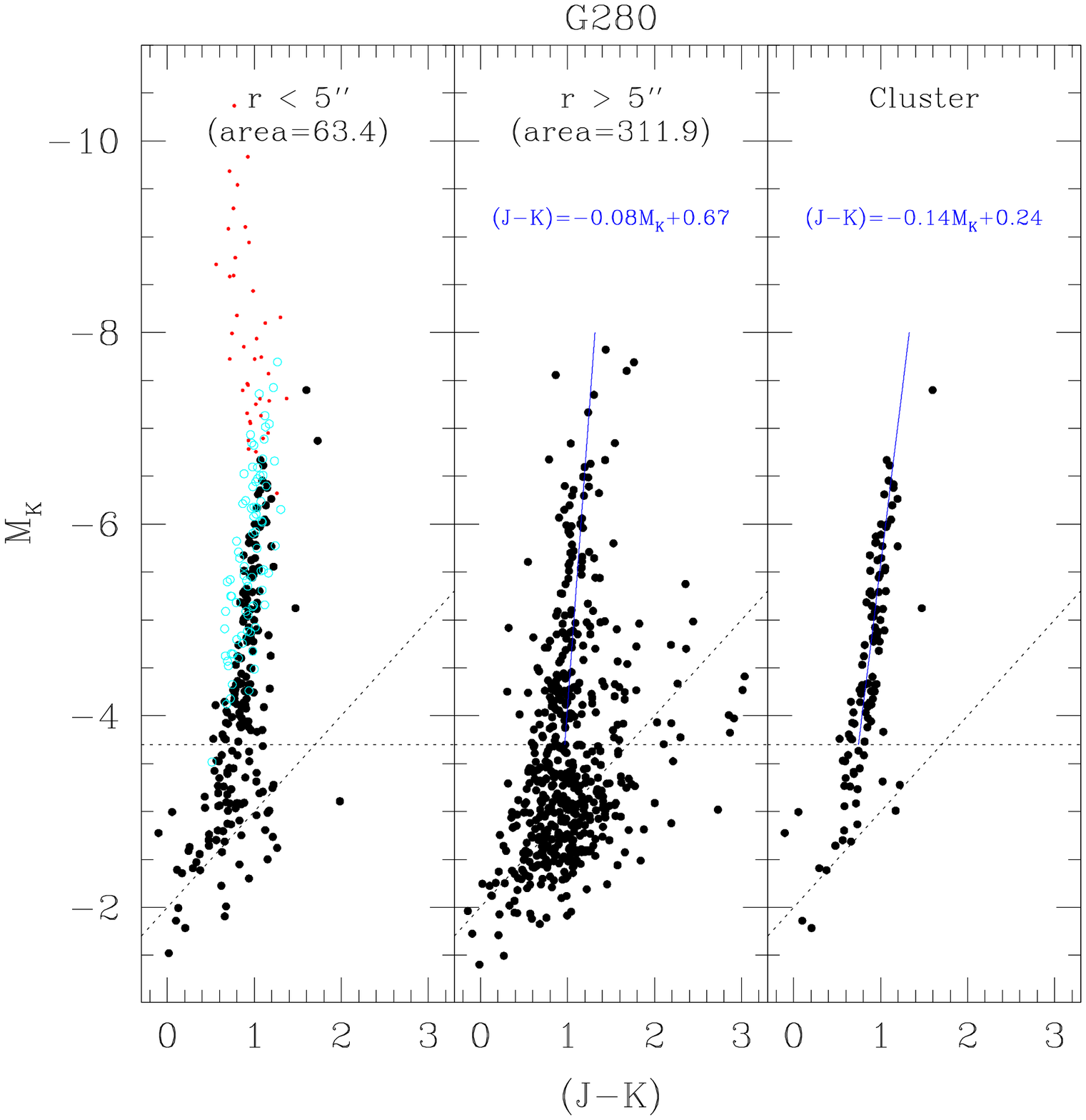}
\figcaption{
G280 -- Left: all objects measured within $5''$ of the cluster center.
Open circles indicate objects within the threshold-blending limit
($\mu_K<16$, $r<2.2''$), and half-size dots are for objects inside the
critical-blending radius ($\mu_K<14$, $r<1.0''$).  Center: objects
farther than $5''$ from the cluster.  Right: statistically field
subtracted cluster GB and the linear fit.  The dashed lines at
$M_K=-3.7$ and $M_J=-2$ indicate the 50\% completeness limits, and all
fits disregard data fainter than these limits.
\label{fig:g280cmd}}
\end{figure}

The G280 $M_K$-$(J-K)$ CMDs are shown in Figure \ref{fig:g280cmd}.  The
left panel shows all the data inside a radius of $5''$, the radius
chosen to define the cluster.  Objects inside the threshold-blending
limit ($\mu_K<16$, $r<2.2''$) are plotted with open circles, and objects
inside the critical-blending limit ($\mu_K<14$, $r<1.0''$) are plotted
with half-size dots.  The center panel shows all objects outside the
$5''$ cluster radius.  These stars are expected to be non-cluster, or
``field'' stars.  The right panel shows the result of statistically
subtracting the field star component from the cluster.  We also omit any
objects we suspect may be affected by blending, so that only objects in
the annulus between the threshold-blending radius $(2.2'')$ and the
cluster radius $(5'')$ are included.

We apply a linear least-squares fit to the cluster stars brighter than
$M_K = -3.7$ and $M_J = -2$, ignoring $3 \sigma$ outliers.  The best-fit
equation is displayed at the top of the right panel.  Using this slope
$(-0.136 \pm 0.011)$ and the GB slope -- [Fe/H] relationship of
\citet{KF1995} we estimate the metallicity of the G280 cluster as $+0.26
\pm 0.36$.

Again, as mentioned in \S \ref{sec:blending_analysis}, the measured
slope, and thus the calculated metallicity, will be affected by
blending.  As in the case of G1, there are so few stars on the upper GB,
that splitting it up into annuli significantly degrades the accuracy of
the slope determination.  We thus take a number -- weighted average
surface brightness of $\mu_K = 16.9$ for the G280 cluster.  This
indicates a metallicity correction of $-0.42 \pm 0.09$ dex is required
to remove the effects of blending.  Applying this correction gives a
final metallicity of $-0.15 \pm 0.37$ for the G280 cluster.

Previous metallicity measurements for G280 are listed in Table
\ref{tab:g1_g280_metallicities}, and the techniques were briefly
discussed in \S \ref{sec:g1}.  These measurements cover a fairly large
range, but our value is consistent with most.

\begin{figure}
\plotone{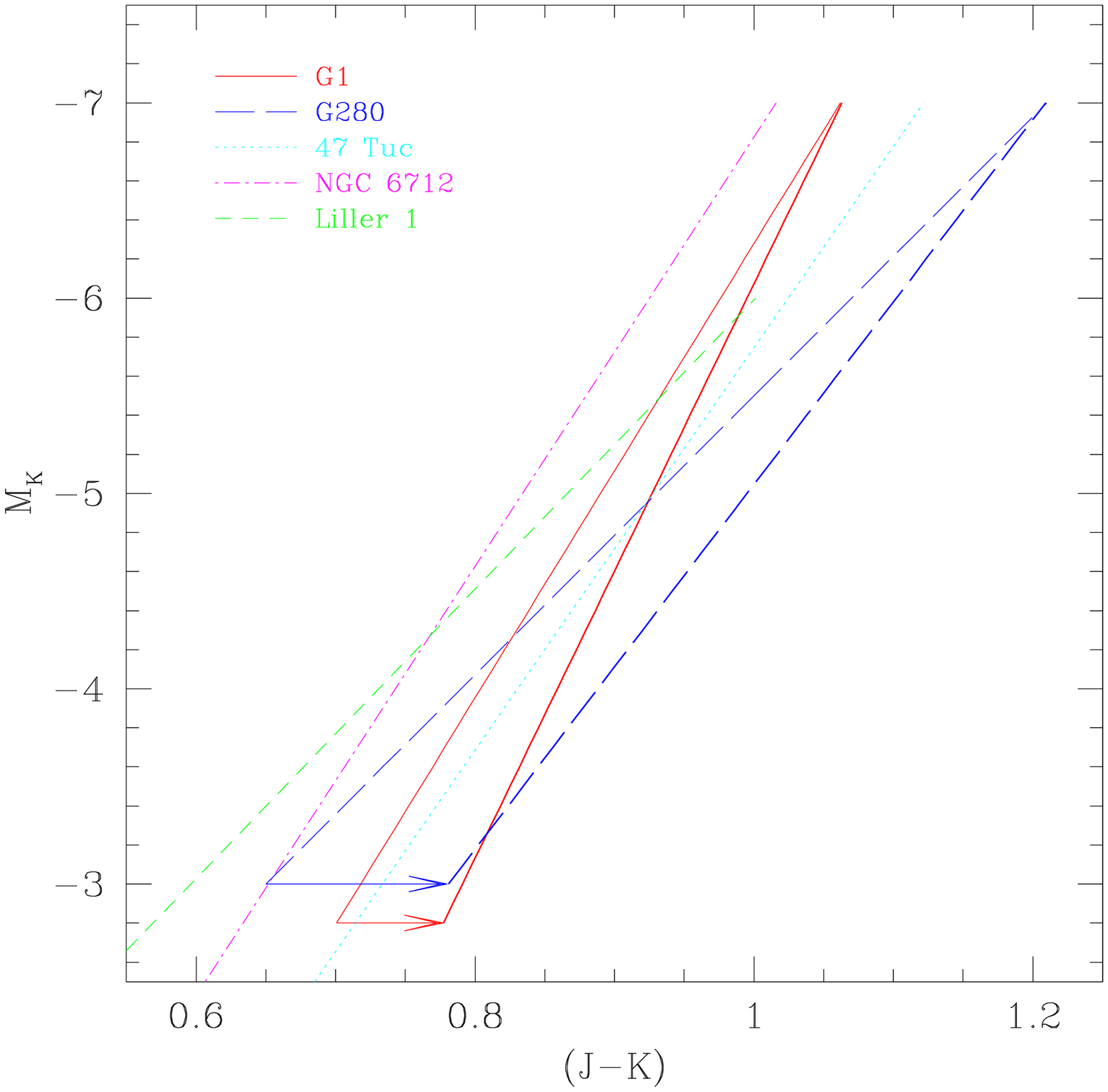}
\figcaption{
Linear fits to the G1 and G280 giant branches.  The arrows show the
shifts estimated for the blending corrections.  Also shown are the GB
fits and metallicities of the Galactic globular clusters 47 Tucanae
[Fe/H]=-0.76, NGC 6712 [Fe/H]$=-1.01$, and Liller 1 [Fe/H]$=-0.2$.
\label{fig:gbs}}
\end{figure}

We also looked for a possible intrinsic color spread, and hence a
metallicity spread in G280.  Such a metallicity spread would not be too
surprising, since G280 is also a very massive cluster, with a velocity
dispersion which is actually higher than that of G1 \citep{DGMC1997}.
However, the dispersion in the measured colors is small, $\sigma_{(J-H)}
= 0.07$ $(-3 > M_H > -6)$, $\sigma_{(J-K)} = 0.08$ $(-4.5 > M_K >
-6.5)$, close to the spread expected solely from measurement errors.
Thus we conclude that, as for G1, G280 does not show any significant
metallicity spread.

Figure \ref{fig:gbs} shows the linear fits to the G1 and G280 giant
branches before and after the blending correction.  Also shown are three
Galactic globular clusters which cover a range in metallicities.  This
figure gives an idea of the relative positions of G1 and G280 in CMD
space, as well as the magnitude and direction of the blending
corrections applied to the GB slopes.

\begin{figure}
\plotone{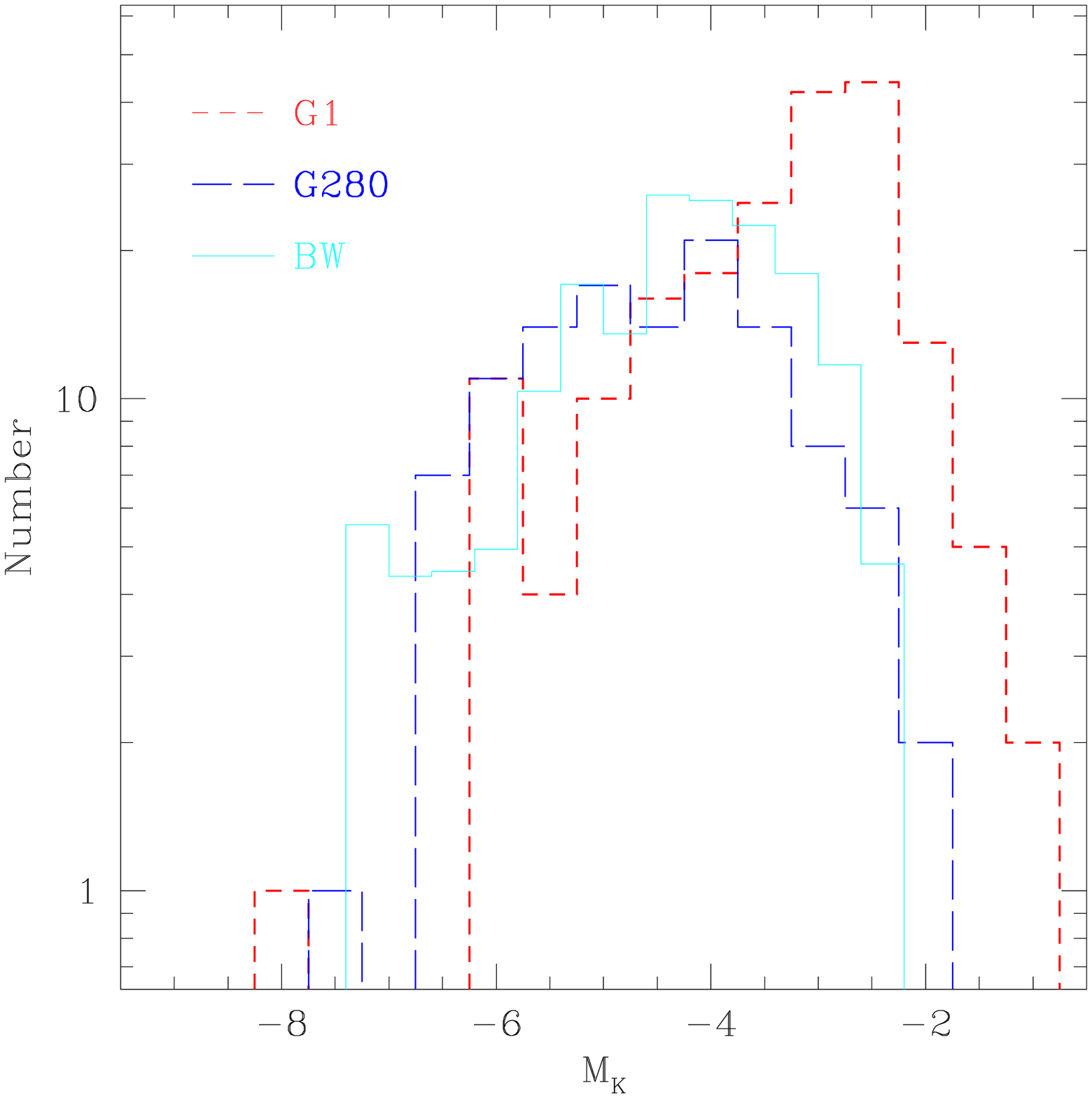}
\figcaption{
The G1 and G280 cluster luminosity functions.  Also shown is the scaled
Baade's Window LF from \citet{FW1987}.
\label{fig:cluster_lfs}}
\end{figure}

The luminosity functions of the G1 (short-dashed) and G280 (long-dashed)
clusters are shown in Figure \ref{fig:cluster_lfs}.  They both appear to
have a sharp cutoff at $M_K \sim -6.5$, which is consistent with what is
observed in Galactic globulars \citep{FMOF2000}.  We also show the
Galactic bulge LF measured by \citet{FW1987} in Baade's Window (solid)
which has a sharp cutoff at $M_K \sim -7.5$.  The faint end of G280
drops off more quickly than that of G1 since it is closer to the nucleus
of M31 and the background is higher, making it more difficult to detect
fainter stars.

%
% Central Clusters
%
\subsection{Central Clusters} \label{sec:central_clusters}

\begin{figure}
\plotone{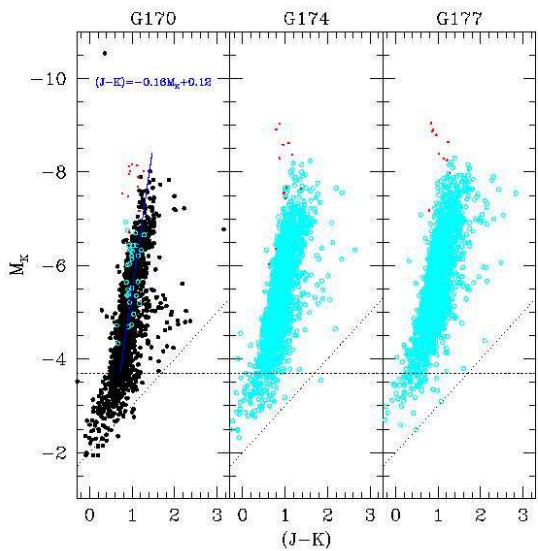}
\figcaption{
CMDs of the fields around G170, G174, and G177.  We provide a linear fit
to the GB of the G170 field, where the crowding is not too severe.  This
fit ignores points inside the critical-blending radius (half-size dots),
points inside the threshold-blending radius (open circles), and objects
fainter than $M_K=-3.7$ or $M_J=-2$.
\label{fig:centralcmds}}
\end{figure}

The three central clusters (G170, G174, \& G177) are so compact that we
were unable to extract any photometry of the cluster stars themselves.
The star counts seem to indicate that the clusters were detected, but
the detections occur in the regime where the effects of blending are
considerable.  Thus for these three clusters we present only the
calibrated field CMDs in Figure \ref{fig:centralcmds}.

We use the same blending criteria as the other clusters, developed in
Paper I.  Objects inside the critical-blending limit $(\mu_K = 14)$ are
plotted with half-size dots, and should be ignored.  (see Table
\ref{tab:blend_radii} for the specific radii of each cluster).  These
are the brightest and bluest stars on the upper RGB of each cluster.
Objects located inside the threshold-blending limit $(\mu_K = 16)$ are
plotted with open circles, and should be considered dubious, note that
nearly all the stars in the G174 and G177 clusters are such dubious
measurements.  We provide a linear fit to the GB measured in the G170
field, but do not attempt to estimate a metallicity, as the effects of
blending are greater than we feel comfortable trying to correct.

As is evident from the radial surface brightness plots in Figure
\ref{fig:surface_brightness}, the stellar density of these central
fields is exceedingly high.  G174 is the closest to the the nucleus of
M31, and G177 is only slightly farther away, but lies along M31's major
axis.  The surface brightness of these two clusters never drops below
the threshold blending limit.  We are therefore skeptical of this
photometry, and do not provide GB fits.  The artificial cluster A177
(the analog of G177), revealed brightening even in the field of up to
0.6 magnitudes at $K$ due to blending.

%
% Surrounding fields
%
\subsection{Surrounding Fields}

\begin{figure}
\plotone{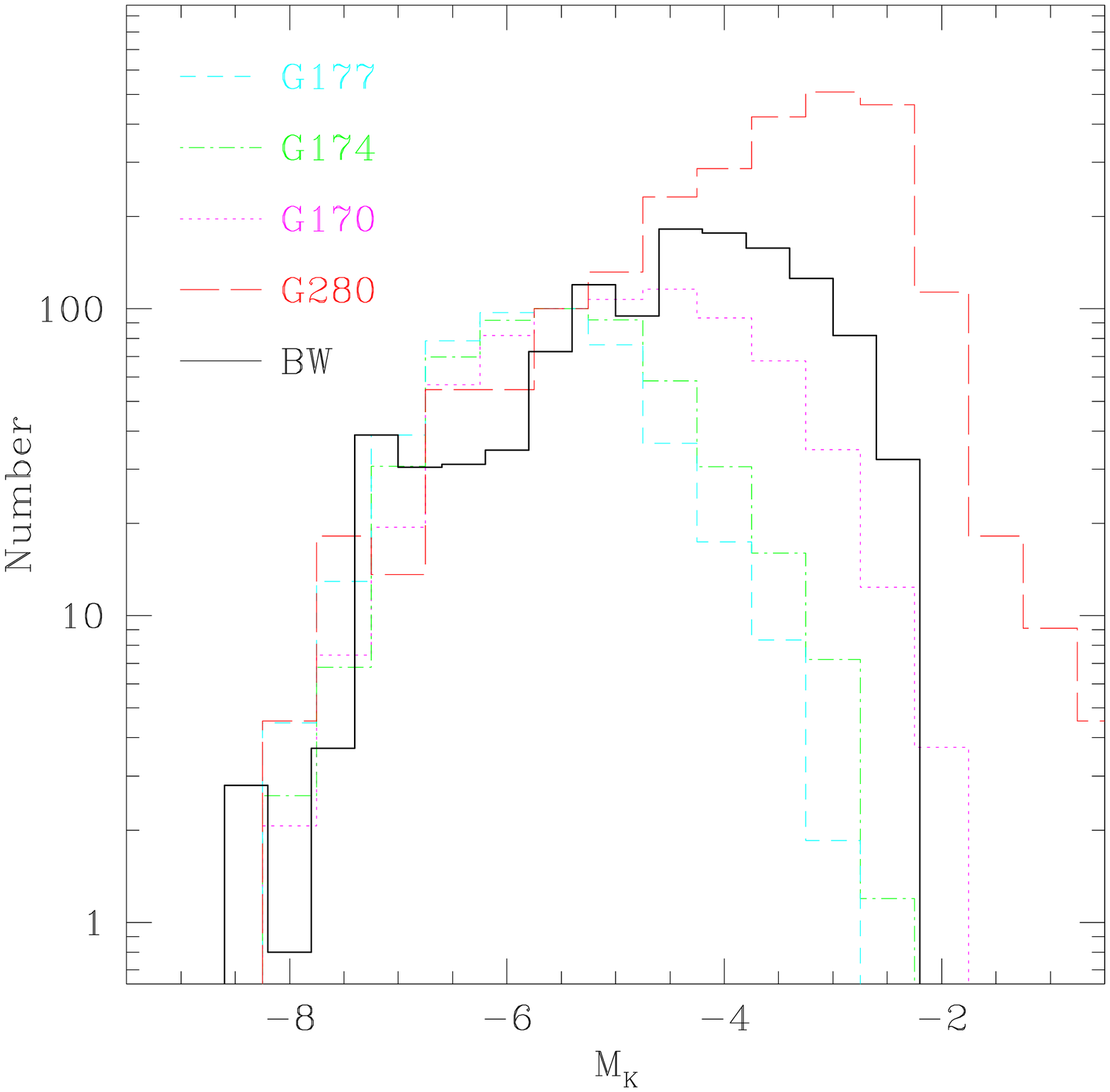}
\figcaption{
Normalized luminosity functions for the fields around each cluster, and
the Baade's Window LF from \citet{FW1987}.  The similarity of these LFs
to the BW LF indicates that there is no significant population of bright
young stars in the bulge of M31.
\label{fig:fieldlfs}}
\end{figure}

Figure \ref{fig:fieldlfs} shows the scaled luminosity functions for the
fields surrounding the cluster observations.  The G1 field is not
included as it is assumed that almost all of the stars are cluster
members since only 2 were found in the nearby control field.  The G280
field includes all stars that are greater than $5''$ from the cluster
center.  For the central clusters, we plot everything outside their
threshold-blending radii, which are listed in Table
\ref{tab:blend_radii}.

All the field LFs look approximately the same, allowing for differences
in the faint end due to varying levels of incompleteness.  They all have
an upper limit of $M_K \sim -8$, similar to what has been measured in
the Galactic bulge where \citet{FW1987} observed a sharp break at $M_K
\sim -7.5$ and stars trailing off to $M_K \sim -8.5$.  These
observations imply that the bulge of M31 has a stellar population not
significantly younger than that in Baade's Window, contrary to some
previous observations.  We will address this issue in more detail when
we present our NICMOS data targeted at M31's bulge

%
% The G280 Field
%
\subsubsection{The G280 Field} \label{sec:g280_field}

Of all the field observations, the G280 field is the least crowded, and
the only one where we feel comfortable performing a giant branch
analysis.  The field CMD is shown in the center panel of Figure
\ref{fig:g280cmd}.  The linear fit to the GB, rejecting points fainter
than $M_K=-3.7$ and $M_J=-2$, as well as $>3 \sigma$ deviants, is shown
at the top of the panel.  The slope of the GB is $-0.081 \pm 0.022$,
which yields a metallicity of $-1.05 \pm 0.58$ from the relation of
\citet{KF1995}.  The average surface brightness of the field is $\mu_K
\sim 18.3$, which requires a metallicity correction of $-0.23 \pm 0.09$
dex to compensate for the effects of blending (see \S
\ref{sec:blending_analysis}).  Thus our final estimate of the
metallicity of the G280 field is $-1.28 \pm 0.59$.

The large error associated with the field metallicity estimate is not
due to measurement errors, but rather to the very large spread in color.
The dispersion in the $(J-H)$ color is $\sigma_{(J-H)} \sim 0.10$, and
the dispersion in the $(J-K)$ color is $\sigma_{(J-K)} \sim 0.14$, both
of which are significantly larger than the measurement errors.  This
spread in color is most likely due to a true spread in metallicity.
Such a spread is not unexpected, since this field samples both the disk
and bulge populations.

\citet{FCP1983} have empirically determined a relationship between
the GB $(J-K)$ color at $M_K=-5.5$ and the cluster metallicity using 12
Galactic globulars with [Fe/H] $< -0.8$.  This relation has been
recently revised by \citet{FMOF2000} using 10 GGCs with more accurate
distances: [Fe/H] $= 4.76 \times (J-K)_0(GB)_{-5.5} - 5.38$.  The
measured spread in $(J-K)$ for the M31-G280 field is $\sigma_{(J-K)}
\sim 0.14$ in the range $-4.5 > M_K > -6.5$.  We estimate the $(J-K)$
spread due to measurement errors as $\sigma_{(J-K)} \sim 0.04$ (also the
spread measured in G1).  Thus the intrinsic spread in color is
$\sigma_{(J-K)} \sim 0.13$.  Using the \citet{FMOF2000} relation, this
leads to a spread in metallicity of $\sigma_{\rm [Fe/H]} \sim 0.6$,

Since our $J$ and $H$ band data are the deepest, we have used the
evolutionary tracks of \citet{GBBC2000} to derive a relation between the
GB $(J-H)$ color at $M_H=-4$ and the cluster metallicity.  We find that:
[Fe/H] $= 4.76 \times (J-H)_0(GB)_{-4.0} - 4.27$.  In the G280 field the
measured spread in $(J-H)$ is $\sigma_{(J-H)} \sim 0.10$ in the range
$-3.0 > M_H > -6.0$.  We estimate the spread in $(J-H)$ due to
measurement errors is $\sigma_{(J-H)} \sim 0.06$ (also the spread
measured in G1).  Thus the intrinsic spread in $(J-H)$ color is
$\sigma_{(J-H)} \sim 0.08$.  Using our relation derived from the
evolutionary tracks, this corresponds to a spread in metallicity of
$\sigma_{\rm [Fe/H]} \sim 0.4$,

Combining these measurements, keeping in mind that the $(J-H)$ data have
higher signal-to-noise, and the $(J-K)$ relation is empirically
determined, we estimate that the true spread in metallicity for the G280
field is $\sigma_{\rm [Fe/H]} \sim 0.5$.  Note that no correction for
blending is required, since our analysis shows no significant blending
effects until brighter than $\mu_K \sim 16$ magnitudes arcsecond$^{-2}$

%
% M31's disk (from the G280 field)
%
\subsubsection{M31's Disk} \label{sec:m31's_disk}

One would expect, that at $20.5'$ from the center of M31, the field
surrounding G280 should have a significant contribution from M31's disk.
However, the luminosity function of this field looks very similar to
those obtained in our fields in the bulge of M31.  In this section, we
estimate the disk contribution to this field, and an age for this disk
component based on the luminosity of the brightest AGB stars.

To find the relative contributions of the disk and bulge, we use the
bulge--disk decomposition of \citet{Ken1989}.  We take the position of
G280 as $20.5'$ from the center of M31, and 34\degr\ from the major
axis.  An interpolation of Kent's data shows that G280's position
corresponds to a major axis distance of $\sim 23.8'$, where the $r$-band
surface brightness is $\mu_r = 21.23$ magnitudes arcsecond$^{-2}$.  The
decomposition reveals that 85\% of the flux is from the disk
($\mu_r(disk)=21.4$), and 15\% is from the bulge ($\mu_r(bulge)=23.3$).
Thus the stellar population of the disk should be well represented in
the G280 field.

To estimate the age of the disk stars in our field, we first convert our
$K$-band measurements to bolometric luminosities using the corrections
of \citet{FPC1980a}.  These corrections use the $(J-K)$ color, and we
apply their M-star correction since this is a relatively high
metallicity field.  The results show that the brightest stars 
are mostly fainter than $M_{bol} \sim -5$.

\begin{figure}
\plotone{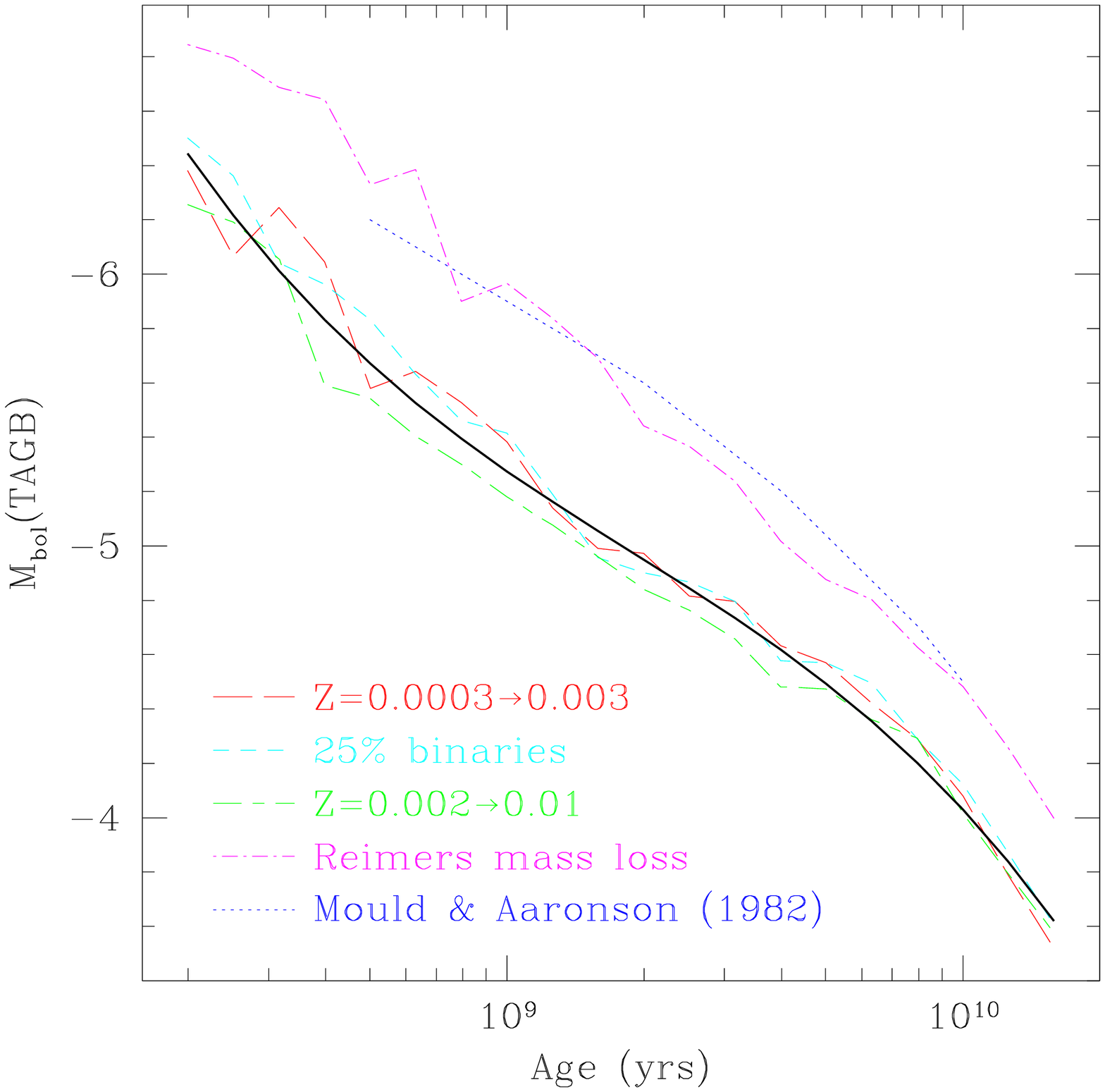}
\figcaption{
Theoretical luminosity of the AGB tip as a function of population age.
Our calculations use constant star formation from 15 Gyr to 7 Myr ago,
metallicity increasing linearly with time, and the \citet{BMCB1992}
stellar evolution models.  The solid line is a polynomial fit to the
three models using the \citet{VW1993} mass loss prescription.
\label{fig:mboltagb}}
\end{figure}

Assuming that these few bright stars are members of a young population,
we can estimate the age of this population using the relationship
between AGB tip luminosity and age.  First used by
\citet{MA1979,MA1980}, this relation makes use of the monotonically
decreasing maximum luminosity of the TAGB with age.  Theoretically this
relation is valid only for populations which are homogeneous in age and
chemical composition, certainly not what we are observing in this field.
However, as we will show, the dependence with metallicity is small (see
Fig \ref{fig:mboltagb}), and we are only trying to estimate the age of
the youngest stars from the brightest observed $M_{bol}$.  Due to the
limited number of stars in our field, and the short lifetime at the
TAGB, this estimate is only an upper limit to the age of the youngest
stars.

We have calculated this relation using the ZVAR synthetic CMD code of
\citet{BMCB1992} \citep[see also][for descriptions]{GABC1996,GFABC1999}
to produce a synthetic CMD with constant SFR from 15 Gyr to 7 Myr ago,
and metallicity increasing linearly with time using the \citet{BMCB1992}
stellar evolution models.  We have used the mass-loss prescription of
\citet{VW1993}, (see \citet{GABC1996} for a discussion on the effects of
different mass loss prescriptions on the AGB morphology of CMDs of
composite stellar populations).  Of all the parameters used, mass loss
is the one that has the strongest effect on the magnitude of the TAGB as
a function of age.

This relation differs from that derived by \citep{MA1982} in which they
use a mass loss treatment as parameterized by \citet{Rei1975} with
$\eta=0.45$.  At any given age, our predicted TAGB is $\sim 0.5$ mag
fainter (Fig \ref{fig:mboltagb}), and as a consequence we are deriving
younger ages for stars of the same apparent magnitude.

Figure \ref{fig:mboltagb} illustrates our results for several different
scenarios.  As previously stated, all cases use constant SFR from 15 Gyr
to 7 Myr ago, metallicity increasing linearly with time and the
\citet{BMCB1992} stellar evolution models.  The first three lines listed
in the legend all use the newer mass loss prescription of
\citet{VW1993}, and show the relative insensitivity of the TAGB
luminosity to binary fraction and metallicity.  The first model (long
dash) has Z increasing from 0.0003 to 0.003 with no binaries.  The
second model (short dash) has the same metallicity, but now with 25\%
binaries.  The third (short/long dash) has Z increasing from 0.002 to
0.01 with no binaries.  We fit these three models with a smooth
polynomial and plot it with a solid line.  It is this fitted relation
($m_{bol}(TAGB)=-401.42+126.16\log(age)-13.492\log(age)^2+0.48499\log(age)^3$)
which we use to estimate the age of the youngest stars in the G280
field.  For comparison we show the luminosity -- age relation of
\citet{MA1982} which uses mass loss from \citet{Rei1975}.  We have also
run our low-metallicity, no binaries model using the Reimers mass loss
prescription, and it falls on the \citet{MA1982} relation.  This
verifies that the primary difference is indeed the treatment of
mass-loss.

Using the relation just developed, we estimate the age of the youngest
stars in the G280 field.  Considering only the single measurement of the
bolometric luminosity of the brightest star in our field of
$M_{bol,m}=-5.0$, we estimate an age of $\sim 2$ Gyr.  One can also
attempt to statistically estimate the tip of the AGB using the technique
of \citet{AM1982}.  In this case one averages the stellar luminosities
brighter than $-4.5$, and assuming that the distribution of stars along
the AGB is uniform, the peak luminosity is twice the mean.  Following
this procedure, we average the (two) stars brighter than $-4.5$ and find
a mean of $\overline{M_{bol}(<-4.5)} = -4.8$.  which leads to an AGB tip
luminosity of $M_{bol,f}=-5.15 \pm 0.28$ for a fully populated AGB.
According to our relation derived from the stellar models, this implies
an age of $1.3^{+1.1}_{-0.6}$ Gyr for the disk stars in this field.
Thus we conclude that the youngest component of this field can be as
young as 1-2 Gyr based on the luminosities of the brightest giants.

Although the measured GB color of the G280 field is very similar to that
of Baade's Window, it is not inconsistent with having a small component
of the population as young as 2 Gyr.  This is because the GB color is
not very sensitive to age.  Our models show that at $M_K=-6$ the $(J-K)$
color only changes by $\sim -0.05$ going from 10Gyr to 2Gyr.  This is
confirmed by the Yale Isochrones using the LeJeune color tables to get
infrared colors.

%
% CONCLUSIONS
%
\section{Conclusions} \label{sec:conclusions}

We first present surface brightness profiles of all clusters, and with
the blending analysis presented in Paper I, determine radial photometric
limits for each cluster.  We then give integrated photometry of all
clusters except G1, which was not fully on the detector.

For the G1 cluster, we present the infrared CMD, and estimate the
metallicity as [Fe/H]$= -1.22\pm0.43$ from the slope of the giant
branch.  Based on the width of the giant branch, which shows no
significant spread in color over what is expected from measurement
errors alone $(\sigma_{(J-H)}=0.06)$, we conclude that there is no
significant metallicity spread in the cluster.  We combine our infrared
observations of G1 with two epochs of optical $V$-band HST-WFPC2 data,
revealing that several of the brightest stars in the cluster are LPVs.
The shape and position of the GB in the $K$-$(V-K)$ CMD are similar to
that of M107, indicating a metallicity of [Fe/H]$=-0.9\pm0.2$.  However
since the infrared GB slope technique uses such a small range in
luminosity, we place more weight on the higher value from the
optical-infrared CMD.

For the G280 observations, we divide the frame into cluster and field at
$5''$ from the cluster center.  We statistically subtract the field
population from the cluster, and present both the cluster and field
CMDs.  Fitting the giant branch, we find a cluster metallicity of
[Fe/H]$= -0.15\pm0.37$.  As in G1, we see no evidence for a metallicity
spread in the cluster based on the width of the GB
$(\sigma_{(J-H)}=0.07)$.

Fitting the GB of the G280 field, we find a metallicity of $-1.3 \pm
0.6$.  The large error on the metallicity is indicative of the large
color spread, which we estimate to be $\sigma_{[Fe/H]} \sim 0.5$ dex
from the width of the GB.  This is not surprising, since this field has
contributions from both the disk (85\%) and bulge (15\%).

What is surprising is that, in this field which is 85\% disk, we see no
obviously bright, young stars.  Using the brightest star in the field as
the tip of the AGB, at $M_{bol}=-5$, we estimate an age of $\sim 5$ Gyr.
However, if the disk component were this young, we would expect to see
$\sim 30$ stars brighter than $M_{bol}=-4$, but we see only 7.  A more
likely scenario is that there are just a few young disk stars in the
field, while the majority of the disk population is closer to $\sim
10-15$ Gyr, thus lowering the AGB tip to $M_{bol} \sim -4.5$.

The three central clusters, G170, G174 \& G177, are all too compact to
extract cluster star photometry.  We thus present the field CMDs and
luminosity functions without trying to separate the cluster and field
star contributions.  The surface brightness of the G170 field is within
acceptable limits, so we perform a linear fit to the GB, but do not try
to estimate the metallicity, as the blending correction would be
uncomfortably large.  The G174 and G177 fields, on the other hand, are
both above the threshold-blending surface brightness limit.  This
implies that measurements of all but the brightest stars in these fields
are potentially affected by blending, so we refrain from even fitting
their GBs.

Finally we presented the cluster (G1 and G280) and field luminosity
functions with the LF measured in Baade's Window.  The luminosity
functions of G1 and G280 both have a sharp bright-end cutoff at $M_K
\sim -6.5$, consistent with observations of Galactic globulars.  The
fields surrounding the clusters have LFs which are indistinguishable
from that measured in the Galactic bulge.  Thus, at least in the fields
observed, there is no significant population of young luminous stars in
the bulge of M31.

\

\acknowledgements
Support for this work was provided by NASA through grant number GO-7826
from the Space Telescope Science Institute.  RMR acknowledges additional
support from NASA contract NAG5-9431, awarded in connection with the
NGST ad hoc science working group.  Thanks to Peter Stetson for
supplying, and helping us with his photometry package, ALLFRAME.
Valuable comments from Darren DePoy, Paul Martini and Marcia Rieke were
greatly appreciated.

% Figure 1
%\clearpage
%\begin{figure}
%\plotone{fig1.eps}
%\figcaption{blah}
%\end{figure}

\end{document}